\documentclass[preprint,prl]{revtex4}

\usepackage{epsfig}

\def\e{\begin{equation}}
\def\f{\end{equation}}
\def\ea{\begin{eqnarray}}
\def\fa{\end{eqnarray}}
\def\=#1{\overline{\overline{#1}}}
\def\_#1{{\bf #1}}

\def\.{\cdot}

\begin{document}

\title{Near-field enhancement and imaging in double
cylindrical polariton-resonant structures: Enlarging perfect lens}

\author{Pekka Alitalo}

\author{Stanislav Maslovski}

\author{Sergei Tretyakov}

\affiliation{Radio Laboratory / SMARAD, Helsinki University of Technology\\
P.O. Box 3000, FI-02015 TKK, Finland\\
{\rm E-mails: pekka.alitalo@tkk.fi, stanislav.maslovski@tkk.fi,
sergei.tretyakov@tkk.fi}}

\date{\today}

\begin{abstract}

We experimentally demonstrate a prototype of a cylindrical {\em
enlarging} lens capable of enhancing and restoring evanescent
fields. The enabling phenomenon is the resonant excitation of
coupled surface modes in a system of two cylindrical arrays of
small resonant particles. As was shown in [J.~Appl.~Phys. {\bf
96}, 1293 (2004)], this phenomenon in planar arrays can be used in
electromagnetic near-field imaging. Here, we use a similar
structure in a cylindrically symmetric configuration, which gives
us a possibility to obtain an enlarged near-field image.

\end{abstract}


\maketitle

\section{Introduction}

Since J. Pendry found that a slab filled by a material with
negative parameters $\varepsilon$ and $\mu$ theoretically operates
as a perfect lens~\cite{Pendry1}, enhancing evanescent components,
many researchers tried to realize such a device experimentally.
The main difficulty in the design is the problem of losses in the
artificial material filling the lens. Recently, we proposed an
alternative approach to enhancement and imaging of evanescent
fields, which is based on two parallel planar lattices of small
resonant particles~\cite{we}. Since no volumetric artificial media
are used in this device, the issue of losses can be mitigated.
Experiments in the microwave region have demonstrated imaging of
near fields in these new structures~\cite{we}. Evanescent fields
are enhanced on lattices of small resonant particles, because
these  planar structures support propagation of slow
electromagnetic waves, which is also the key property of
interfaces between vacuum and materials with negative parameters.
Such waves are known also as surface or Zenneck waves in radio
science, or as surface plasmons or polaritons in optics. These
waves are called slow because their phase velocity (along the
surface) is less than the speed of light. Respectively, the fields
of such waves decay exponentially from the surface.

The ideal planar perfect lens restores  the field distribution of
the source (both amplitude and phase) in the image plane. A
natural extension would be the use of more complex lens shapes to
enable also enlargement of the image. Using geometrical optics, it
has been theoretically shown that image enlargement is indeed
possible in a perfect lens of a circular cylindrical
shape~\cite{Pendry2}. However, no attempts to realize this effect
using volumetric metamaterials experimentally have been made.

Our purpose here is to experimentally show enhancement of
evanescent fields and enlarged images of source field distribution
using arrays of resonant particles. For this purpose we construct
two concentric lattices of particles, creating a cylindrically
symmetrical system. We restrict ourselves by the case when the
surface modes propagate around the cylinders, i.e. we consider
only the angular spatial spectrum. Contrary to the infinite planar
geometries this spectrum is discrete.

To excite slow waves, either a non-uniformity in the surface is
needed, or the incident field must be in phase synchronism with
these waves. The latter means that the exciting field must be an
evanescent plane wave. When the propagation factor (along the
surface) of the incident wave coincides with the propagation
factor of a surface wave, a strong resonance occurs. In this case
the surface wave is most effectively excited.

The source of evanescent field and the cylindrical arrays are
placed in a model cell described in~\cite{we}. Essentially, this
cell is a planar waveguide operating below cut-off. Because of the
symmetry of the waveguide modes, a circular ring of particles
placed inside the waveguide behaves as an infinitely long cylinder
under incidence of an evanescent wave. We will show that the
surface modes excited on the arrays can be used to amplify near
fields of a source and, respectively, to obtain an image mimicking
the distribution of the source field on a larger scale.

\section{Experiment}

\begin{figure}[ht]
\centering \epsfig{file=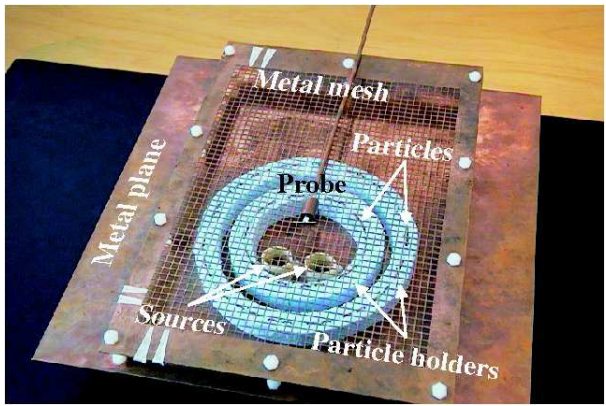, width=8cm}
\caption{(Color online). Experimental setup used to measure the
distribution of the evanescent fields inside the waveguide.}
\label{experimental_setup}
\end{figure}

The experimental setup was similar to that used in~\cite{we}. The
setup consisted of a two-plate waveguide, in which the distance
between the metal plates ($h$) was smaller than $\lambda/2$
($\lambda$ is the wavelength in free space). The setup is shown in
Fig.~\ref{experimental_setup}. The evanescent source fields were
excited by two horizontal $\lambda/4$-dipoles placed inside the
waveguide. The operating frequency of the setup was close to 5 GHz
and therefore $h$ was chosen to be 2.5 cm ($\lambda$ at 5 GHz is 6
cm).

In order to measure the fields inside the waveguide, the upper
plate of the waveguide was made of a dense metal wire mesh, which
was weakly penetrable to the fields inside the waveguide (the
transmission coefficient of the mesh was measured to be about
$-20$ dB for normal plane-wave incidence at 5 GHz). The
distribution of electric field inside the waveguide could then be
measured using a probe, which was a horizontal $\lambda/4$-dipole
placed approximately 3 mm above the metal mesh and controlled by a
robot connected to a PC.

The source dipoles were connected to port 1 of a vector network
analyzer (VNA) using a power divider and the probe was connected
to port 2 of the VNA. By scanning the surface of the metal mesh
with the probe and at the same time measuring $S_{21}$ with the
VNA, it was possible to measure the relative field distribution
(in the horizontal plane) inside the waveguide.

\begin{figure}[ht]
\centering \epsfig{file=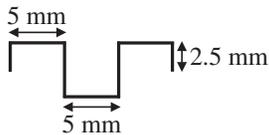, width=3.5cm}
\caption{A resonant particle with $f_r\approx5.464$ GHz. Diameter
of the copper wire is 0.8 mm.} \label{resonant_particle}
\end{figure}

The cylindrical arrays consisted of small particles. These
particles, in which the surface resonance occurs, were made of
thin copper wires that were meandered to make the overall particle
size considerably smaller than the resonant wavelength, see
Fig.~\ref{resonant_particle}. First a large amount of these
resonant particles was manufactured and each was tested to have
approximately the same resonant frequency ($f_r\approx5.464$ GHz).
The testing was done by placing the particle under test inside the
waveguide and close to a source dipole (distance between the
particle and the source dipole was approximately 2 cm). When the
field magnitude was measured as described before, the field on top
of the resonant particle had a sharp peak (much stronger than the
field on top of the source dipole) at a certain frequency. This
frequency was then determined to be approximately the same for all
the resonant particles used.

First an array consisting of ten resonant particles and a radius
of $r=40$ mm was constructed. The array was placed inside the
waveguide around two source dipoles (the dipoles were
approximately in the center between the two plates of the
waveguide) with the other end of the array about 15 mm from the
dipoles. The resonant particles were placed in a dielectric foam
holder ($\varepsilon_r\approx 1$), so that the particles were
approximately in the center between the waveguide's two plates.

\begin{figure}[ht]
\centering \epsfig{file=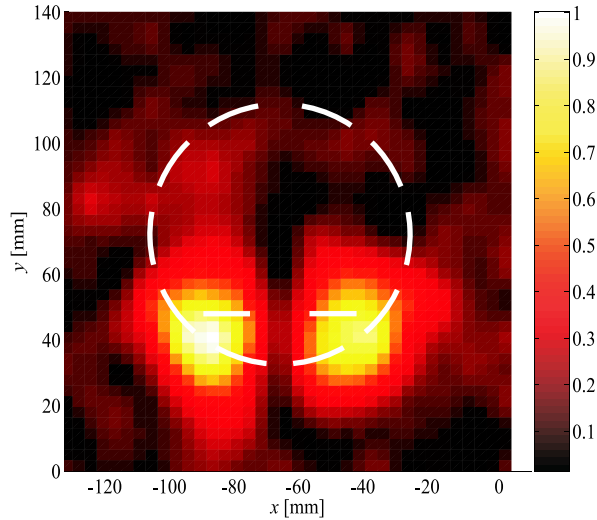, width=8cm}
\caption{(Color online). Distribution of electric field amplitude
measured on top of the waveguide ($f=5.28$ GHz). The probe dipole
is oriented along the $x$-axis. Two evanescent sources are placed
inside a cylindrical array made of resonant particles. The places
of the sources and the array are shown with white color.}
\label{small array}
\end{figure}

When the fields were measured, two strong field maxima occurring
at the measuring frequency of $f\approx 5.28$ GHz were observed on
top of the array, see Fig.~\ref{small array}. The fields in
Fig.~\ref{small array} were measured with the probe dipole
oriented along the $x$-axis. By measuring the same situation with
the probe oriented along the $y$-axis, it was seen that the field
distribution was approximately the same as shown in
Fig.~\ref{small array} (the two maxima were at the same places and
no other maxima were seen). As can be seen from Fig.~\ref{small
array}, the array was strongly excited near the two sources. Also
another resonant frequency was found ($f\approx 5.37$ GHz) at
which also the other side of the array was excited (symmetrically
with that shown in Fig.~\ref{small array}, i.e. four maxima
altogether).

Next, another, larger array consisting of 18 resonant particles
arranged at a circle with the radius $r=70$ mm was manufactured.
The measurements showed that this array had many resonant
frequencies, with one of them most pronounced. By measuring the
field distribution as before (with the other side of the array
close to the two sources, as in Fig.~\ref{small array}), it was
noticed that there existed a resonance near 5.3 GHz. By manually
shifting the positions of the particles, the resonant frequency of
the array could be tuned to coincide with the resonant frequency
of the smaller array. The array was tuned to resonate at $f=5.28$
GHz, corresponding to the resonant frequency at which the fields
in Fig.~\ref{small array} were measured.

\begin{figure}[ht]
\centering \epsfig{file=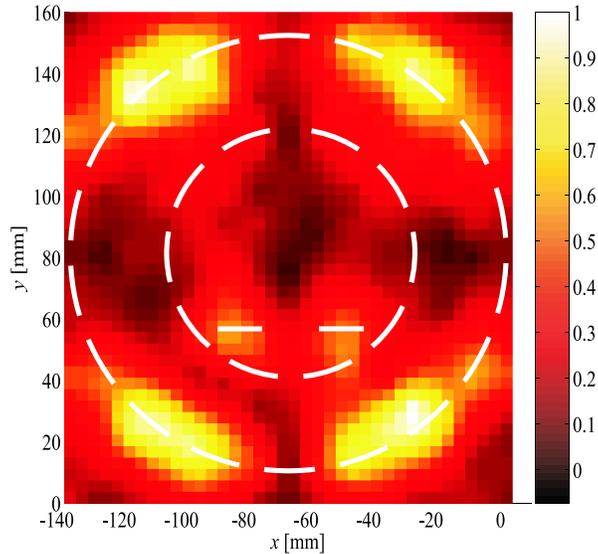, width=8cm} \caption{(Color
online). Distribution of electric field amplitude measured on top
of the waveguide ($f=5.28$ GHz). The probe dipole is oriented
along the $x$-axis. Two evanescent sources are placed inside two
coaxial cylindrical arrays made of resonant particles. The places
of the sources and the arrays are shown with white color.}
\label{two_arrays}
\end{figure}

The particles of the larger array were excited only near the two
sources when measured at this frequency. Again two maxima were
seen on top of the array. Next, both arrays were placed around the
sources in a coaxial manner, see Fig.~\ref{two_arrays}. The
geometry of this configuration is the same as for a volumetric
perfect lens theoretically studied in~\cite{Pendry2}. From the
field plot it is seen that the smaller array was weakly excited
and the larger array had four strong field maxima (images of the
source field), as shown in Fig.~\ref{two_arrays}. Similar results
were obtained with planar arrays in~\cite{we}, where the resonant
array situated closer to the source was not excited at all, but
the resonant array situated further had a very strong maximum.

The maxima of the field at the image side appear to follow the
same angular distribution as that of the source field. The two
images on the other side of the larger array can be considered as
``parasitic'' images. The fields in Fig.~\ref{two_arrays} were
measured with the probe dipole oriented along the $x$-axis. By
measuring the same situation with the probe oriented along the
$y$-axis, it was seen that the field distribution was
approximately the same as shown in Fig.~\ref{two_arrays} (the four
maxima were at the same places and no other maxima were seen).

The larger array was also tuned to match the other resonant
frequency of the smaller array (i.e. $f=5.37$ GHz). Using that
frequency, the two ``parasitic'' images seen in
Fig.~\ref{two_arrays} did not appear.

\section{Conclusion}

We have experimentally demonstrated a prototype of an enlarging
perfect lens, using amplification of evanescent fields on two
cylindrical arrays of small particles. The arrays supported
surface waves at certain frequencies. By tuning the arrays to
resonate at the same frequency, the evanescent source fields
inside the two arrays were amplified and reached the maximum
values on the outer array. According to the measured field
distributions, the angular distribution of the source field
located inside the two coaxial arrays is properly restored on the
outer array. The angular resolution of the image is mainly
determined by the period of the arrays. Resonant meandered pieces
of copper wires were successfully used in this microwave concept
demonstration. For higher-frequency realizations other small
resonating particles may be used, such as plasmonic nano-spheres
for potential optical applications.

\end{document}